\documentclass[pre,twocolumn,twoside,latterpaper,showpacs,floatfix]{revtex4} 
\usepackage{graphicx}   
\usepackage{epsfig}
\usepackage{latexsym}

\begin{document} 
\title{Headgroup dimerization in methanethiol monolayers on the Au(111) surface: a density functional theory study } 
\author{Jian-Ge Zhou, Quinton L. Williams, Frank Hagelberg}   
\affiliation{Computational Center for Molecular Structure and Interactions, 
Department of Physics, Atmospheric Sciences, and Geoscience, Jackson 
State University, Jackson, MS 39217, USA }  
\author{}   
\affiliation{} 
  
\begin{abstract}
A long-standing controversy related to the dimer pattern formed
by S atoms in methanethiol ($CH_{3}SH$) on the Au(111) 
surface has been resolved using density functional 
theory. For the first time, dimerization of methanethiol 
adsorbates on the Au(111) surface is established by computational modeling.
For methylthiolate ($CH_{3}S$), it is shown that the S atoms do not dimerize 
at high coverage but reveal a dimer pattern at intermediate coverage. 
Molecular dynamics simulation at high coverage demonstrates that the observed dialkyl 
disulfide species are formed during the desorption process, and thus are not attached to the surface.

\end{abstract}  
\maketitle   

\section{Introduction}

Recently, much attention has focused on the properties of 
self-assembled monolayers (SAM), particularly of alkanethiol on gold surfaces \cite{fee}-\cite{rgs}; 
see Ref. \cite{sv} for a review.
This high level of interest may be ascribed to their relevance to wetting phenomena, tribology, 
chemical and biological sensing, optics and nanotechnology. 
Despite the apparent simplicity of these systems, their observation in various experiments has led to 
contradictory results, in particular related to the question if the S-H headgroups of alkanethiol organize in dimers on 
the Au(111) surface or not. 

In the early phase of SAM related studies, it was assumed that alkanethiol adsorbs at
threefold coordinated hollow sites on the Au(111) surface \cite{sv}. This picture has been first challenged by 
grazing incidence X-ray diffraction (GIXRD) studies \cite{fee,fsb}, where a dimer pattern was proposed. 
High-resolution electron-energy-loss spectra supported this dimer model \cite{kcm}. Scanned-energy and 
scanned-angle photoelectron diffraction experiments on alkanethiolate, in contrast, did not yield any 
dimerization \cite{kis,skn}. Based on temperature-programmed desorption, Auger electron
spectroscopy, and scanning tunneling microscopy (STM) studies, methanethiol was found to form dimers \cite{msd}. 
Recently, Torrelles et al. 
observed by use of GIXRD, STM and
electrochemical techniques that alkanethiolate does not dimerize \cite{tvv}. 

The present contribution addresses these controversial observations by simulation from first principles.
A consistent model is proposed that agrees with all available experimental data related to thiol 
adsorption on the Au(111) surface. 
Until recently, it was assumed that the S-H bond of methanethiol is cleaved once it is 
adsorbed to the Au(111) surface, such that it becomes methylthiolate \cite{gca,sv}. Thus,
methanethiol and methylthiolate would exhibit the same adsorption pattern once deposited on Au(111).
Recent results, however, show that methanethiol stays intact when in contact with the 
defect-free Au(111) surface \cite{rlm,zh}, and turns into methylthiolate
on Au(111) only in the presence of vacancies. To examine if methanethiol dimerizes on Au(111), 
it is crucial to study the methanethiol and methylthiolate dimer patterns separately.
If these two species display different adsorption 
structures, the long-standing controversy related to their dimerization behavior might be resolved.

Guided by this motivation, we study in the present contribution the adsorption of methanethiol 
and methylthiolate on the Au(111) 
surface by use of density functional theory \cite{khf}.
First, we will present adsorption energies and geometries for methanethiol on the 
regular Au(111) at 0.5 ML and 1.0 ML, 
and demonstrate, for the first time, that methanethiol reorganization on the clean Au(111) surface
gives rise to the formation of a dimer pattern. 
Subsequently, we will show that methylthiolate on the regular Au(111) surface 
exhibits a coverage 
dependent dimerization behavior. 
Further, we address the possible impact of surface defects and perform molecular dynamics (MD) 
simulations to inspect the effect of temperature on the dimer formation. 

\section{Computational Method}

\begin{figure}
\includegraphics[width=3.5in]{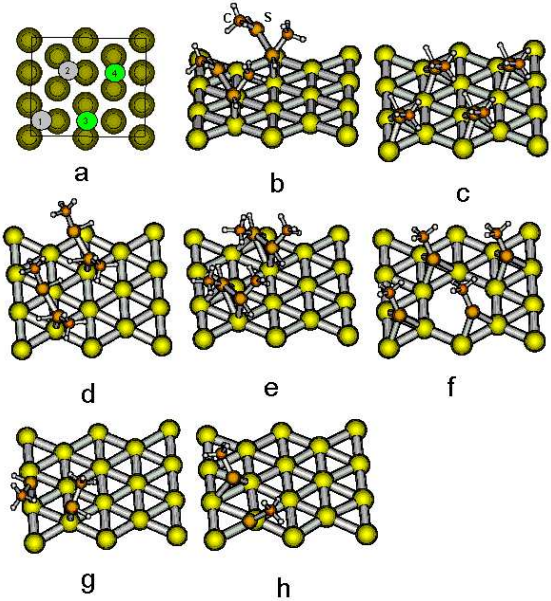}
\caption{(Color online) (a) $(3\times 2\sqrt{3})$  supercell. 
(b) top+hcp. (c) hcp-hcp. (d) top+hcp'. (e) top+fcc+3b.
(f) hcp-hcp$^*$ with a vacancy. (g) top+top. (h) top+fcc+3b'. }  
\label{4x2}
\end{figure}

We employ the VASP code \cite{khf} which involves the projector
augmented wave potential \cite{kj,peb} 
and the PW91 generalized gradient approximation \cite{pw}.
The wave functions are expanded in a plane wave basis with an energy cutoff of 400 eV. The Brillouin zone integration 
is performed by use of the Monkhorst-Pack scheme\cite{mp}. 
We utilized a $3\times 3\times 1$  k point mesh for the geometry optimization.
A $(3\times 2\sqrt{3})$ superlattice (see Fig. \ref{4x2}a) is employed as an Au(111) supercell \cite{vgs}, 
corresponding to
12 Au atoms per layer. The Au atoms in the 
top three atomic layers are allowed to relax, 
while those in the bottom layer are fixed to simulate 
bulk-like termination \cite{zhx}. 
The vacuum region comprises seven atomic layers, which
exceeds substantially the extension of the methanethiol molecule.
To examine the accuracy of our approach we increased the energy cutoff to 500 eV and the number of 
k points to $8\times 8\times 1$. In both cases, the difference amounted
to less than $2\%$. We further calculated the gold lattice constant, 
and found it to agree with the experimental value \cite{ksu} within  $2.1\%$. 

\section{Results and Discussion}
\subsection{Methanethiol molecules deposited on the regular Au(111) surface}
In what follows we present our results, beginning with a discussion of both the geometries and adsorption
energies for the optimized configurations of methanethiol on the regular Au(111) surface at 
0.5 ML (Fig. \ref{4x2}g, Fig. \ref{4x2}h) and 
1.0 ML (Fig. \ref{4x2}b - Fig. \ref{4x2}f) coverage, as displayed in Table \ref{thiol}. 
At 1.0 ML, we arrange four methanethiol molecules in the 
$(3\times 2\sqrt{3})$  supercell according to the experimentally detected structure \cite{ccl,bsk}.
The molecules labeled
1(3) and 2(4) are symmetry equivalent, see Fig. \ref{4x2}a. 
\begin{table}
\caption{The geometries and adsorption energies for the studied methanethiol configurations 
on the regular Au(111) surface at 0.5 ML and 1.0 ML. The entries
$d_{S-S}$, $d_{S-Au}$ (in $\AA$) and $E_{ads}$ (in eV/per
methanethiol) refer to the S-S bond length, 
the shortest Au-S bond length and the adsorption energy, respectively. 
The superscripts $I$ and $E$ stand for the initial and equilibrium structures, respectively. 
The maximum adsorption energy is underlined.}
\begin{center}
\begin{tabular}{cc|ccc|ccc}  
\hline
  ~&~&~ &0.5ML&~&~&1.0ML&~\\
Initial & $d_{S-S}^{I}$ & $d_{S-S}^{E}$ & $d_{S-Au}^{E}$ & $E_{ads}$ &  $d_{S-S}^{E}$ &  $d_{S-Au}^{E}$ & $E_{ads}$\\
\hline
bri+fcc&2.30&3.36 &2.69&0.46&3.76 &3.02 &0.23\\
bri+fcc'&2.30&3.69&3.05&0.37&3.74 &2.99 &0.23\\
bri+hcp&2.30&3.51&2.86&0.42&4.23 &2.83 &0.25\\
bri+hcp'&2.30&3.78&2.93&0.38&3.74 &2.98 &0.23\\
bri-bri&3.90&4.20&2.93&0.24&4.01 &2.73 &0.18\\
bri-top&3.90&3.65&2.89&0.31&3.66 &3.21 &0.15\\
fcc+hcp&2.30&4.47&2.79&0.46&3.07 &2.63 &0.19 \\
fcc+hcp'&2.30&3.73&2.71&0.40&3.73 &2.45 &-1.99$\#$\\
fcc-bri&3.70&4.35 &2.88&0.28&3.95 &3.03 &0.18\\
fcc-fcc&5.00&5.08 &3.15&0.24&5.08 &3.67 &0.15\\
fcc-top&3.40&3.88&3.17&0.29&3.83 &3.21 &0.14 \\
hcp+fcc&2.30&4.28&2.77&0.44&3.10 &2.69 &0.21\\
hcp+fcc'&2.30&3.62&2.73&0.40&3.06 &2.47 &-0.70$\#$\\
hcp-bri&4.20&4.23&3.17&0.24&4.24 &3.12 &0.18\\
hcp-fcc&3.40&4.11 &3.11&0.23&4.00 &3.29 &0.18\\
hcp-hcp&5.00&5.08&3.16&0.23&5.08 &3.60 &0.14\\
hcp-hcp$^*$&5.00&5.08&3.08&0.25&5.08 &3.33 &0.15\\
hcp-top&4.50&4.96&3.05&0.29&4.26 &3.47 &0.21\\
top+fcc&2.30&3.50&2.65&0.40&3.15 &2.57 &0.20\\
top+fcc+3b&2.30&3.43&2.68&0.46&3.25 &2.80 &0.20\\
top+fcc'&2.30&3.21&2.68&0.37&3.74 &3.26 &-0.59$\#$\\
top+fcc+3b'&2.30&3.21&2.76&0.31&3.61 &2.48 &-2.03$\#$\\
top+hcp&2.30&3.48&2.68&0.39&3.54 &2.64 &$\underline{0.26}$\\
top+hcp+3b&2.30&3.48&2.65&0.43&3.85 &3.14 &0.25\\
top+hcp'&2.30&3.26&2.67&0.39&3.27 &2.82 &0.25\\
top+hcp+3b'&2.30&3.27&2.79&0.32&3.29 &2.85 &0.21\\
top+top&2.30 &3.58 &2.70 &$\underline{0.48}$ &3.36 &2.88 &0.23\\
top-top&5.00&5.08&3.03&0.36&5.08 &3.17 &0.19\\
top-top'&5.90&5.96&3.04&0.35&3.67 &2.77 &0.20\\
\hline
\end{tabular}
\end{center}
\label{thiol}
\end{table}
The nomenclature top+hcp in Table \ref{thiol} refers to an initial geometry 
where the S atom of adsorbate 1
is placed on an exact top site, while that of adsorbate 3 is just in an fcc 
region so that 
the distance between the S atoms of 1 and 3 is close to 2.3 $\AA$ (see Fig. \ref{4x2}b), 
and analogously for the notations bri+hcp, hcp+fcc, etc. In Fig. \ref{4x2},
the S and C atoms are distinguished by ball symbols of different sizes.
The symbol hcp-hcp in Table \ref{thiol} indicates that the S atoms of both, adsorbates 1
and 3, are placed exactly at a hcp center too
(Fig. \ref{4x2}c), with analogous definitions for the 
symbols bri-bri, top-top etc. The arrangements top+hcp'  (Fig. \ref{4x2}d) and top+hcp differ by the 
dihedral angle of the two S-C bonds, and accordingly for other primed structures. The top+fcc+3b configuration is shown
in Fig. \ref{4x2}e, where the S-S group forms three bonds with the surface.

From Table \ref{thiol}, at 0.5 ML the adsorption energy for the most stable structure is 0.48 eV
(top+top, see Fig. \ref{4x2}g), and the S adsorption sites are
on the top of Au atoms, in keeping with recent experimental results \cite{msd}. The S-S distance
is about 3.6  $\AA$. 
Thus at 0.5 ML coverage a dimer is formed
when methanethiol is adsorbed to the regular Au(111) surface at low temperature. 
This dimerization was indeed observed recently by STM experiment at 5 K \cite{msd}. 
In Tables \ref{thiol}, \ref{thiolate} and \ref{defect}, some configurations at 1.0 ML coverage are marked 
by $\#$. These structures 
are not favored since they involve broken S-C bonds. From Table \ref{thiol}, 
the adsorption energy of the stable configuration at 1.0 ML, with top+hcp as initial 
structure, is 0.26 eV. 
The optimized structure involves one S atom in the fcc region but tending toward the atop site; the other 
S atom locates in the fcc region, leaning toward the bridge site, in agreement with experimental findings \cite{rlm}. 
The resulting S-S distance is about 3.6 $\AA$. This implies methanethiol dimer formation also at 1.0 ML coverage. 

The bond between the methanethiol and Au(111) involves an ionic contribution due to electron transfer 
proceeding from the surface to the adsorbate. This interaction is well described by
the density functional theory (DFT) \cite{psb}. 
Our calculations yield an effective charge of -0.2 e, per methanethiol molecule.
The binding energy of two pure methanethiol molecules, maintaining their spatial 
relation as realized
on the surface, results as 0.03 eV by DFT \cite{khf}, 
0.02 eV using a coupled cluster with single, double and triple excitations (CCSD(T)/6-311+G(d,p)) \cite{gau}, and the Van der Waals 
interaction between them is 0.06 eV \cite{gau}.  
Being higher than the CCSD(T) value by 50$\%$, the DFT prediction for the interaction
between two pure molecules is inaccurate, 
which is consistent with recent observations \cite{zt,mg}. 
However, the interaction that leads to dimerization is a surface mediated effect, which strongly exceeds the Van 
der Waals attraction between the dimer constituents \cite{cr}. 
To evaluate the surface mediated interaction between two
dimer constituents, we study the case of 0.5 ML where the preferred configuration is 
the optimized top+top structure (Fig. \ref{4x2}g).
Specifically, we consider the difference $\Delta{E}= E_{ad}^{(top+top)}(d_{S-S}=3.58)-E_{ad}^{(top-top)}(d_{S-S}=5.08)$,
which is interpreted as a reorganization energy, namely the energy gain due to dimer formation by rearranging an initially even adsorbate distribution on the surface.  
This energy is 0.24 eV, and thus twelve times (0.24 eV/0.02 eV(CCSD(T))) larger than the weak 
interaction between two pure methanethiol molecules, maintaining their spatial relation as adopted
on the surface. On the other hand,
the interaction 
energy among two molecules and Au(111) is found to be 0.96 eV (top+top).
The Van der Waals component is thus estimated to contribute only 6.3$\%$ (0.06 eV/0.96 eV) to the total interaction energy. Therefore, the methanethiol dimer formation is 
not caused by the Van der Waals force. Rather, it is a surface mediated effect. 
Since the energy difference between the dimer structures at a S - S distance of 3.6 $\AA$\cite{fee,fsb} and the reference situation of methanethiol adsorbates spread evenly over the Au(111) surface ($d_{S-S}$ around 5.0 $\AA$) \cite{sv} is minimally 0.24 eV at all coverage levels, for both methanethiol and methylthiolate, omission of the relatively small Van der Waals contribution does not change our conclusions regarding the dimerization effect.

\subsection{Methylthiolate adsorbates on the regular Au(111) surface}
Does methylthiolate exhibit the same dimerization behavior as methanethiol? 
Table \ref{thiolate} lists the configurations of methylthiolate on clean Au(111) at two levels of coverage, 
0.5 ML and 1.0 ML,
with the S-S distances close to 2.1 $\AA$, 3.6 $\AA$ and  5.0 $\AA$.
The maximum adsorption energies for 0.5 ML and 1.0 ML, underlined in Table \ref{thiolate}, 
are 2.12 eV and 1.99 eV (per molecule). 
Table \ref{thiolate} shows that
at 0.5 ML, the S-S distance is about 3.6 $\AA$, which is indicative of a dimer (see Fig. \ref{4x2}h). 
At 1.0 ML, an S-S distance of 5.0 $\AA$ is found. In the light of this result, the absence of dimerization reported 
in \cite{kis} appears plausible. The respective measurement involved a coverage level of 
1.0 ML, and from our simulation, no dimers emerge in this case. The lack of dimerization at 1.0 ML for 
methyltiolate as opposed to methanethiol is rationalized by the much stronger substrate-adsorbate interaction 
for the former than for the latter species, as shown by the entries for $E_{ads}$ and $d_{S-Au}$ in
Tables \ref{thiol} and \ref{thiolate}. 

\begin{table}
\caption{The geometries and adsorption energies for methylthiolate
adsorbed on the regular Au(111) surface at 0.5ML and 1.0ML coverage.}
\begin{center}
\begin{tabular}{cc|ccc|ccc}
\hline
  ~&~&~ &0.5ML&~&~&1.0ML&~\\
Initial & $d_{S-S}^{I}$ & $d_{S-S}^{E}$ &    $d_{S-Au}^{E}$ & $E_{ads}$ &  $d_{S-S}^{E}$ &    $d_{S-Au}^{E}$ & $E_{ads}$\\
\hline
bri+fcc&2.30&2.07 &2.70&2.01&2.05 &3.07 &1.82\\
bri+fcc'&2.30&3.43&2.49&1.84&3.16 &2.41 &1.68\\
bri+hcp&2.30&2.07&2.97&1.99&2.06 &2.86 &1.80\\
bri+hcp'&2.30&3.54&2.41&1.88&3.48 &2.42 &1.72\\
bri-bri&3.90&4.33&2.47&1.98&4.40 &2.48 &1.89\\
bri-top&3.90&4.09&2.38&1.92&4.14 &2.39 &1.86\\
fcc+hcp&2.30&2.07&2.61&2.00&2.06 &2.71 &1.87 \\
fcc+hcp'&2.30&3.34&2.40&1.90&3.28 &2.48 &1.35$\#$\\
fcc-bri&3.70&4.31 &2.43&1.92&4.37 &2.42 &1.81\\
fcc-fcc&5.00&5.08 &2.45&2.08&5.08 &2.48 &$\underline{1.99}$\\
fcc-top&3.40&4.15&2.39&1.91&4.89 &2.46 &1.97 \\
hcp+fcc&2.30&2.07&2.58&1.98&2.06 &2.69 &1.86\\
hcp+fcc'&2.30&3.31&2.40&1.83&3.54 &2.45 &1.39$\#$\\
hcp-bri&4.20&3.88&2.44&1.91&3.83 &2.44 &1.91\\
hcp-fcc&3.40&3.83 &2.45&1.90&3.85 &2.44 &1.78\\
hcp-hcp&5.00&5.08&2.47&2.04&5.08 &2.49 &1.96\\
hcp-hcp'&5.00&5.08&2.47&2.10&5.08 &2.47 &$\underline{1.99}$\\
hcp-top&4.50&4.50&2.39&1.88&4.80 &2.50 &1.80\\
top+fcc&2.30&2.06&2.72&2.00&2.05 &2.81 &1.88\\
top+fcc+3b&2.30&2.07&2.78&2.00&2.07 &2.92 &1.87\\
top+fcc'&2.30&2.05&2.83&2.00&3.59 &2.50 &1.40$\#$\\
top+fcc+3b'&2.30&3.66&2.47&$\underline{2.12}$&3.61 &2.48 &-0.51$\#$\\
top+hcp&2.30&2.06&2.75&2.01&2.06 &2.63 &1.86\\
top+hcp+3b&2.30&2.07&2.80&2.00&2.05 &3.08 &1.80\\
top+hcp'&2.30&2.06&2.79&2.00&2.05 &2.86 &1.87\\
top+hcp+3b'&2.30&3.59&2.47&2.04&3.75 &2.48 &1.85\\
top+top&2.30&2.07&2.73&2.00&2.07 &2.78 &1.87\\
top-top&5.00&5.08&2.38&1.73&5.08 &2.38 &1.67\\
top-top'&5.90&5.84&2.38&1.71&2.12 &2.65 &1.84\\
\hline
\end{tabular}
\end{center}
\label{thiolate}
\end{table}

\subsection{Methylthiolate adsorbates on the defected Au(111) surface}
To examine if substrate defects induce S-S dimer formation, 
we allow for a vacancy in the top layer of the $(3\times 2\sqrt{3})$  supercell \cite{vgs} (see Fig. \ref{4x2}f). 
The location of this vacancy is chosen such that it affects as many methylthiolate adsorbates as possible.
It is sufficient to include methylthiolate in this segment of our study 
since methanethiol 
reduces to methylthiolate when adsorbed to the defected Au(111) surface \cite{rlm,zh}.
Table \ref{defect} comprises the methylthiolate configurations in the presence of a 
vacancy at coverage levels of 0.5 and 1.0 ML.

\begin{table}
\caption{The geometries and adsorption energies for methylthiolate
adsorbed on the Au(111) surface with a vacancy.}
\begin{center}
\begin{tabular}{cc|ccc|ccc}
\hline
  ~&~&~ &0.5ML&~&~&1.0ML&~\\
Initial& $d_{S-S}^{I}$ & $d_{S-S}^{E}$ &$d_{S-Au}^{E}$ & $E_{ads}$ &  $d_{S-S}^{E}$ &    $d_{S-Au}^{E}$ & $E_{ads}$\\
\hline
bri+fcc&2.30&2.07&2.55&2.00&2.06 &2.75 &1.85\\
bri+fcc'&2.30&3.25&2.44&2.43&3.30 &2.42 &2.04\\
bri+hcp&2.30&2.06&3.10&2.01&2.06 &2.77 &1.83\\
bri+hcp'&2.30&3.31&2.44&2.43&3.98 &2.44 &1.99\\
bri-bri&3.90&4.22&2.42&2.25&4.44 &2.44 &2.14\\
bri-top&3.90&4.53&2.37&2.13&4.25 &2.38 &2.13\\
fcc+hcp&2.30&2.06&2.53&2.01&2.06 &2.70 &1.87 \\
fcc+hcp'&2.30&4.22&2.46&2.26&3.75 &2.44 &1.53$\#$\\
fcc-bri&3.70&5.00&2.42&2.32&4.19 &2.46 &1.99\\
fcc-fcc&5.00&5.23 &2.37&2.04&4.92 &2.39 &1.99\\
fcc-top&3.40&4.56&2.37&2.12&4.75 &2.44 &$\underline{2.20}$ \\
hcp+fcc&2.30&2.07&2.54&2.08&2.07 &2.55 &1.95\\
hcp+fcc'&2.30&3.81&2.42&2.33&3.62 &2.45 &1.61$\#$\\
hcp-bri&4.20&4.55&2.42&2.37&3.87 &2.42 &2.12\\
hcp-fcc&3.40&4.51 &2.43&2.40&3.78 &2.41 &2.14\\
hcp-hcp&5.00&5.16&2.36&1.98&4.97 &2.44 &2.17\\
hcp-hcp'&5.00&4.73&2.45&2.21&5.03 &2.43 &2.20\\
hcp-top&4.50&4.27&2.36&2.17&4.44 &2.39 &1.92\\
top+fcc&2.30&2.06&2.59&2.08&2.06 &2.68 &1.93\\
top+fcc+3b&2.30&2.07&2.60&2.01&2.07 &2.71 &1.95\\
top+fcc'&2.30&2.06&2.67&2.04&3.67 &2.36 &1.49$\#$\\
top+fcc+3b'&2.30&3.33&2.43&$\underline{2.48}$&3.36 &2.42 &1.61$\#$\\
top+hcp&2.30&2.06&2.62&2.08&2.07 &2.59 &1.89\\
top+hcp+3b&2.30&2.07&2.59&2.02&2.08 &2.57 &1.89\\
top+hcp'&2.30&2.06&2.67&2.04&2.06 &2.66 &1.95\\
top+hcp+3b'&2.30&3.36&2.42&2.45&3.29 &2.49 &2.12\\
top+top&2.30&2.07&2.61&2.09&2.07 &2.64 &1.94\\
top-top&5.00&4.94&2.36&1.87&4.86 &2.43 &2.16\\
top-top'&5.90&5.85&2.37&1.86&2.14 &2.40 &1.96\\
\hline
\end{tabular}
\end{center}
\label{defect}
\end{table}
With respect to dimer formation, the same observations are made as for the regular
surface. The dimerization behavior of methylthiolate is therefore not impacted by the presence
of defects.

\subsection{Molecular Dynamical Simulation}
 In the following, we inspect the effect of temperature on methylthiolate dimerization.
Two methylthiolate molecules are adsorbed on the top layer of one-half of 
the $(3\times 2\sqrt{3})$ 
 supercell, the resulting coverage being 1.0 ML.
\begin{figure}
\includegraphics[width=3.3in]{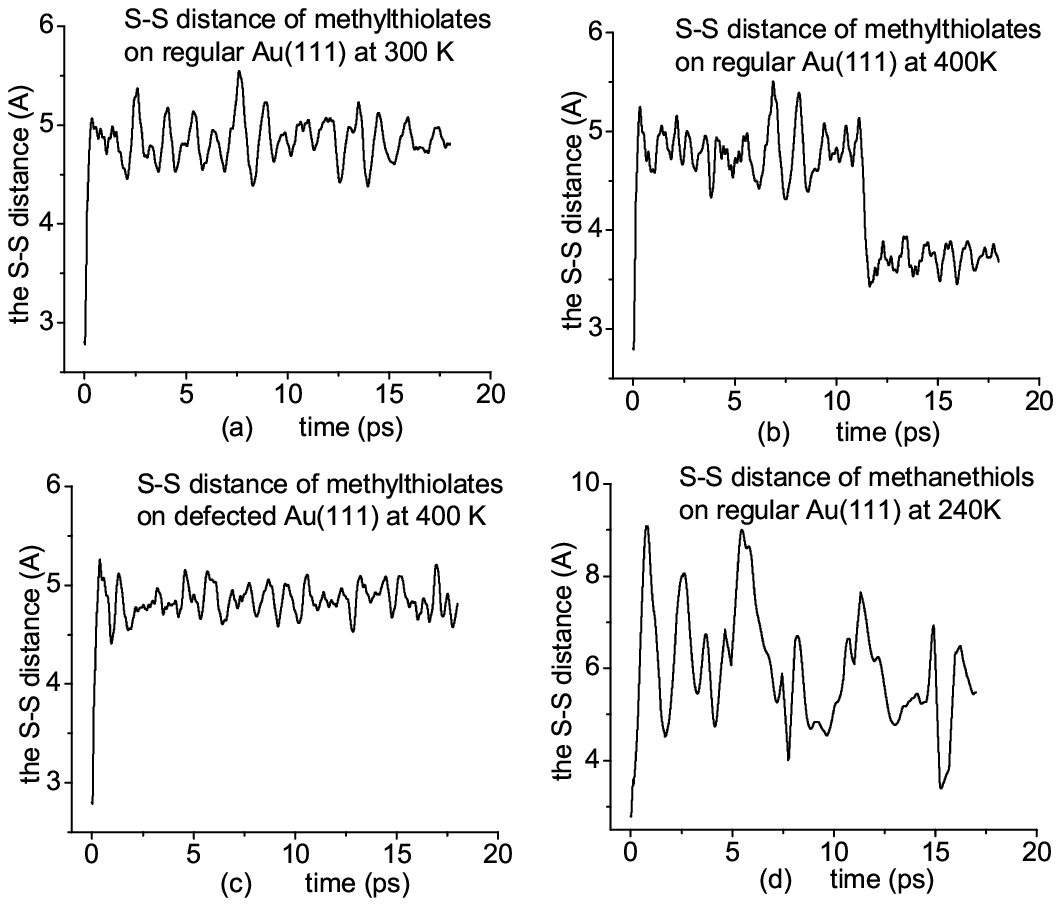}
\caption{Time variation of the S-S 
distance in four cases.}  
\label{md}
\end{figure} 
Fig. \ref{md} shows the time variation of the S-S distance as resulting from MD
 simulation for a) methylthiolate on the regular Au(111) at
300K, b) at 400K, c) methylthiolate on the imperfect Au(111) surface at 400K, d)
methanethiol on the regular Au(111) surface at 240K. At 300K, methylthiolate is still attached to the regular 
Au(111) surface, the S-S 
distance oscillates around 5.0 $\AA$, which shows that the adsorbates do not dimerize at this temperature. As one raises the temperature to 400K, however, the
S-S distance reduces to 3.5 $\AA$. The temperature increase weakens the molecule-surface bond and thus favors intermolecular bonding. This process leads to dimer formation. If the temperature is high enough and methylthiolate 
starts to desorb, two molecules join to form dimethyl disulfide (CH$_{3}$SSCH$_{3}$). 
Thus the observed dialkyl disulfide emerges during the 
desorption process, and does not exist on the surface \cite{rlm,kcm}. At 400K, methylthiolate forms a 
dimer on regular Au(111), while the S-S distance turns out to be 5.0 $\AA$ for 
imperfect Au(111), and no dimerization is found (Fig. \ref{md}(c)). 
In the presence of a vacancy, the interaction between methylthiolate and Au(111) 
is stronger than in the regular case \cite{mh,zh,mcv}, and interaction between the molecules is accordingly weaker. For methanethiol, no dimers exist at 240 K.
 
\section{Summary}

In conclusion, we have demonstrated for the first time that at low temperature, 
the methanethiol S atoms form dimers when adsorbed on the regular Au(111) surface \cite{rlm}.
As the temperature rises, these dimers dissociate. 
The S atoms in methylthiolate do not dimerize at 1.0 ML, 
while a dimer emerges as most stable at 0.5 ML. With increasing temperature the 
S-S distance of two adjacent methylthiolate molecules shortens, 
leading to the formation of the experimentally observed dialkyl disulfide. This chemical dimer emerges during the 
desorption process and does not exist on the surface. Prior to this study, it was assumed
that methanethiol and methylthiolate realize the same pattern upon adsorption on the Au(111) surface. 
This assumption was based on the premise that methanethiol deposited on a gold substrate turns into methylthiolate.
This proposition, however, has turned out to be fallacious. As demonstrated by both experiment and theory \cite{rlm,zh},
methanethiol persists on the clean Au(111) surface, and this work establishes that 
methanethiol and methylthiolate display distinctly different dimerization patterns.
Our results thus resolve the long-standing controversy over the correct 
interpretation of the experimental data related to the dimer formation by 
methanethiol and methylthiolate headgroups on the Au(111) surface. 
\section*{Acknowledgments}
This work is supported by the NSF through Grants HRD-9805465 and DMR-0304036, 
by NIH through Grant S06-GM008047, by AHPCRC under Cooperative Agreement 
No. DAAD 19-01-2-0014, and by DoD through Contract $\#$W912HZ-06-C-005.

\end{document}